\documentclass[sn-mathphys,Numbered]{sn-jnl}

\def\jsm#1{{\bf #1}}
\usepackage{graphicx}%
\usepackage{multirow}%
\usepackage{amsmath,amssymb,amsfonts}%
\usepackage{amsthm}%
\usepackage{mathrsfs}%
\usepackage[title]{appendix}%
\usepackage{xcolor}%
\usepackage{textcomp}%
\usepackage{manyfoot}%
\usepackage{booktabs}%
\usepackage[ruled,linesnumbered]{algorithm2e}
\usepackage{algorithmicx}%
\usepackage{algpseudocode}%
\usepackage{listings}%

\usepackage{bm}
\usepackage{multicol}

\begin{document}

\title[Article Title]{Noise suppression in photon-counting CT using unsupervised Poisson flow generative models}


\author*[1,2]{\fnm{Dennis} \sur{Hein}}\email{dhein@kth.se}

\author[3,4]{\fnm{Staffan} \sur{Holmin}}\email{staffan.holmin@ki.se}

\author[5]{\fnm{Timothy} \sur{Szczykutowicz}}\email{TSzczykutowicz@uwhealth.org}

\author[6]{\fnm{Jonathan S} \sur{Maltz}}\email{Jonathan.Maltz@ge.com}

\author[1,2]{\fnm{Mats} \sur{Danielsson}}\email{md@mi.physics.kth.se}

\author[7]{\fnm{Ge} \sur{Wang}}\email{wangg6@rpi.edu}

\author[1,2]{\fnm{Mats} \sur{Persson}}\email{mats.persson@mi.physics.kth.se}

\affil[1]{\orgdiv{Department of Physics}, \orgname{KTH Royal Institute of Technology}, \orgaddress{\city{Stockholm}, \country{Sweden}}}

\affil[2]{\orgdiv{MedTechLabs}, \orgname{Karolinska University Hospital}, \orgaddress{\city{Stockholm}, \country{Sweden}}}

\affil[3]{\orgdiv{Department of Clinical Neuroscience}, \orgname{Karolinska Insitutet}, \orgaddress{\city{Stockholm}, \country{Sweden}}}

\affil[4]{\orgdiv{Department of Neuroradiology}, \orgname{Karolinska University Hospital}, \orgaddress{\city{Stockholm}, \country{Sweden}}}

\affil[5]{\orgdiv{Department of Radiology}, \orgname{University of Wisconsin School of Medicine and Public Health}, \orgaddress{\city{Madison}, \state{WI}, \country{USA}}}

\affil[6]{\orgname{GE HealthCare}, \orgaddress{\country{USA}}}

\affil[7]{\orgdiv{Department of Biomedical Engineering, School of Engineering, Biomedical Imaging Center, Center for Biotechnology and Interdisciplinary Studies}, \orgname{Rensselaer Polytechnic Institute}, \orgaddress{\city{Troy}, \state{NY}, \country{USA}}}

\abstract{Deep learning has proven to be important for CT image denoising. However, such models are usually trained under supervision, requiring paired data that may be difficult to obtain in practice. Diffusion models offer unsupervised means of solving a wide range of inverse problems via posterior sampling. In particular, using the estimated unconditional score function of the prior distribution, obtained via unsupervised learning, one can sample from the desired posterior via hijacking and regularization. However, due to the iterative solvers used, the number of function evaluations (NFE) required may be orders of magnitudes larger than for single-step samplers. In this paper, we present a novel image denoising technique for photon-counting CT by extending the unsupervised approach to inverse problem solving to the case of Poisson flow generative models (PFGM)++. By hijacking and regularizing the sampling process we obtain a single-step sampler, that is NFE=1. Our proposed method incorporates posterior sampling using diffusion models as a special case. We demonstrate that the added robustness afforded by the PFGM++ framework yields significant performance gains. Our results indicate competitive performance compared to popular supervised, including state-of-the-art diffusion-style models with NFE=1 (consistency models), unsupervised, and non-deep learning-based image denoising techniques, on clinical low-dose CT data and clinical images from a prototype photon-counting CT system developed by GE HealthCare.}

\keywords{Deep learning, photon-counting CT, denoising, diffusion models, Poisson flow generative models.}

\maketitle

\section{Introduction}\label{sec1}
X-ray computed tomography (CT) is a medical imaging modality utilized for both diagnosis and treatment planning of a wide range of diseases, including stroke, cancer and cardiovascular disease. However, due to the potential risk posed by even low doses of ionizing radiation, a lot of effort has been put into enabling high diagnostic quality while maintaining the dose as low as reasonably achievable \cite{wang2020,koetzier2023}. Photon-counting CT (PCCT), based on the latest generation of CT detector technology, is able to reduce dose via photon energy weighting and by largely eliminating the effects of electronic noise. PCCT can also enable imaging with higher spatial resolution and produce single-exposure energy-resolved images \cite{willemink2018,flohr2020,danielsson2021,higashigaito2022}. However, obtaining high resolution in either space or energy decreases the number of photons in each respective voxel or energy bin, which increases image noise. Thus, excellent noise performance is required, possibly exceeding the capabilities of today’s state-of-the-art denoising methods.

Deep learning (DL) methods have demonstrated remarkable success in recent years for low-dose and photon-counting CT image denoising\cite{chen2017, wolterink2017, yang2018, kim2019, shan2019, kim2020, yuan2020, li2021, wang2023, niu2023, liu2023, tivnan2023,hein2023}. However, these methods are commonly trained using supervised learning, requiring paired data which is not always available in practice. In particular, paired and perfectly registered clinical images are difficult to obtain, and methods for generating paired data based on simulating low-dose scanning \cite{yu2012} or adding noise maps from phantom scans \cite{huber2023} may be sensitive to imperfect modeling of the systems or to mismatch between patient and phantom geometry. In PCCT, effects such as pulse pileup affect high-dose and low-dose scans differently, further confounding such training schemes. For these reasons, unsupervised and self-supervised methods have become more common \cite{kim2020, yuan2020, wang2023, niu2023, liu2023}.

Diffusion-style models, such as diffusion and Poisson flow models, have demonstrated considerable success for unconditional \cite{sohl-dickstein2015,ho2020,nichol2021,song2021,song2022,karras2022,xu2022,xu2023} and conditional image generation \cite{batzolis2021,chung2022,song2021,saharia2022,saharia2022b}. Notably for medical imaging, this family of models lend themselves very well to inverse problem solving via posterior sampling and have already been demonstrated on a range of inverse problems \cite{song2022b, liu2023, chung2023, tivnan2023, hein2023}. In particular, for the case of diffusion models, it is possible to manipulate the sampling process and re-task the network, trained in an unsupervised manner, to inverse problem solving \cite{song2022b,chung2022,chung2023,liu2023}. This is usually done in two steps. First, information from the prior distribution, the ground truth data, is obtained by estimating its time-dependent score function via denoising score matching, exactly as one does when training an unconditional generator. Once equipped with the estimated score function, one can draw samples from the desired posterior distribution by augmenting the sampling processing with a data consistency step that regularizes the generative process, forcing the sample to be consistent with the input \jsm{(conditioning)} image. Moreover, it is unnecessary to run the sampling process all the way from an initial sample from the prior noise distribution. Indeed, it is even beneficial to ``hijack'' the sampling process by inserting a version of the condition image at some stage of the reverse diffusion \cite{chung2022,chung2023}. Hijacking will not only help regularize the problem further, but it will also result in faster sampling;  a smaller number of function evaluations (NFE) being required to achieve the desired image quality. Note that this is a very efficient strategy, since the learned score function can successfully be used to solve a range of different inverse problems, or downstream tasks, without retraining \cite{song2022b,chung2023}. It is also possible to directly learn to solve the inverse problem via supervised learning \cite{tivnan2023, hein2023}. However, a supervised approach requires paired data. 

The NFE required for diffusion-style models, such as diffusion and Poisson flow models, may be on the order $10^1-10^3$ for both conditional (image-to-image) and unconditional (noise-to-image) generation. This may limit their usage in applications where speed is critical, such as clinical CT image denoising. Efforts to reduce the NFE required include moving to efficient ordinary differential equation (ODE) samplers \cite{song2022} and distillation techniques \cite{salimans2022}. Consistency models \cite{song2023} build on the probability flow ODE formulation of diffusion models and learn a consistency network that maps any point on the trajectory to its initial point, including the final point which is just a sample from the prior noise distribution. As such, it achieves single-step sampling (NFE=1). A consistency model may be distilled from a pre-trained diffusion model using so-called consistency distillation (CD) or it can be obtained as a stand-alone model using consistency training (CT).   

In this paper, we present a novel image denoising technique for low-dose and photon-counting CT that extends posterior sampling Poisson flow generative models (PPFM) \cite{hein2023} to the case when paired data are not available. PPFM is a diffusion-style posterior sampling image denoising method that exploits the added robustness of the Poisson flow generative models (PFGM)++ framework to enable NFE=1 whilst maintaining high image quality. In particular, we first train an unconditional PFGM++ for the task of image generation on randomly extracted patches from the prior, ground truth, images. The sampling process is subsequently hijacked and regularized, as in PPFM, to enforce consistency with the input, condition, image. The main contributions of this paper are as follows: 1) We present unsupervised PPFM, a novel image denoising technique that extends PPFM \cite{hein2023} to the case when paired data is unavailable. 2) We demonstrate that it is possible to efficiently train the network, in an unsupervised manner, on extracted patches from the ground truth data and subsequently manipulate the sampling process to denoise full resolution images. Training on randomly extracted patches is more efficient in terms of graphics memory requirements and provides additional regularization. 3) The proposed method contains a posterior sampling diffusion model 
(EDM \cite{karras2022}) as a special case when $D\rightarrow \infty$ and our results indicate that the added flexibility of choosing $D$ as a hyperparameter results in improved performance over having $D \rightarrow \infty$ as is the case for diffusion models. 4) We evaluate the proposed method on clinical low-dose CT images and clinical images from a prototype PCCT system developed by GE HealthCare (Waukesha, WI) \cite{almqvist2023}. We demonstrate that the proposed method performs competitively compared to current state-of-the-art diffusion-style models with NFE=1, consistency models \cite{song2023}. Notably, the consistency model is trained in a supervised fashion, whereas our proposed method is unsupervised. Despite imposing a significantly laxer data requirement, the proposed method performs on par quantitatively, and better qualitatively. In addition to consistency models, we also compare our proposed method to popular supervised, unsupervised, and non-deep learning-based image denoising techniques.

Code used for this project is available at: \url{https://github.com/dennishein/pfgmpp_PCCT_denoising}.

\section{Methods}\label{sec2}
\subsection{Diffusion models and PFGM++}
Diffusion models (EDM \cite{karras2022}) and PFGM++ \cite{xu2023} both work by iteratively denoising images following some physically meaningful trajectory --- inspired by non-equilibrium thermodynamics and electrostatics, respectively. Despite the widely different underlying physics, diffusion models and PFGM++ are intimately connected in theory and in practice. The training and sampling processes of PFGM++ converge to those for diffusion models in the $D\rightarrow \infty, r=\sigma \sqrt{D}$ limit \cite{xu2023}. Thus, diffusion models are incorporated as a special case of PFGM++. In addition, it is possible to reuse the training and sampling algorithms in EDM \cite{karras2022} for PFGM++ by a simple change of variables and an updated prior noise distribution \cite{xu2023}. Expanding on the probability flow ODE formulation in \cite{song2021}, EDM \cite{karras2022} writes the ODE as
\begin{equation}
    d\bm{x} = - \dot{\sigma}(t)\, \sigma(t)\,\nabla_{\bm{x}}\log p_{\sigma(t)} (\bm{x}) \, dt, 
    \label{edm_ode}
\end{equation}
where $\nabla_{\bm{x}}\log p_{\sigma(t)} (\bm{x})$ is the time-dependent score function of the perturbed distribution, and $\sigma(t)$ noise scales. This ODE defines a trajectory between an easy-to-sample prior noise distribution, a simple Gaussian, and the data distribution of interest. Intuitively, running the ODE forward or backward in time respectively nudges the sample toward or away from, the prior noise distribution. Notably, Eq.\  \eqref{edm_ode} depends only on the data distribution via the gradient of the log-likelihood, also known as the score function. Let $p(\bm{y})$ represent the data distribution, $p(\sigma)$ the distribution of noise scales, and $p_{\sigma}(\bm{x}|\bm{y})=\mathcal{N}(\bm{y},\sigma^2\bm{I})$ the Gaussian perturbation kernel. One can then estimate the score function via the perturbation based objective
\begin{equation}
    \mathbb{E}_{\sigma \sim p(\sigma)} \mathbb{E}_{\bm{y}\sim p(\bm{y
    })} \mathbb{E}_{\bm{x}\sim p_{\sigma}(\bm{x}|\bm{y})} \notag 
    \left[ \lambda(\sigma) ||f_{\theta}(\bm{x},\sigma)-\nabla_{\bm{x}} \log p_{\sigma}(\bm{x}|\bm{y})||_2^2\right], 
\end{equation}
where $\lambda(\sigma)$ is a weighting function. Given our estimate of the time-dependent score function, we can generate a sample by solving Eq. \eqref{edm_ode}, using a iterative ODE solver, starting from an initial sample from the prior noise distribution. 

PFGM++ \cite{xu2023} works by treating the $N$-dimensional data as electric charges in a $(N+D)$-dimensional augmented space. Let $\tilde{\bm{y}}:=(\bm{y},\bm{0}) \in \mathbb{R}^{N+D}$ and $\tilde{\bm{x}}:=(\bm{x},\bm{z}) \in \mathbb{R}^{N+D}$ denote the augmented ground truth and perturbed data, respectively. The object of interest in then the high dimensional electric field 
\begin{equation}
    \bm{E} (\bm{\tilde{x}}) = \frac{1}{S_{N+D-1}(1)} \int \frac{\bm{\tilde{x}}-\bm{\tilde{y}}}{||\bm{\tilde{x}}-\bm{\tilde{y}}||^{N+D}} \, p(\bm{y}) \, d\bm{y},
    \label{pfgmpp_E}
\end{equation}
where $S_{N+D-1}(1)$ is the surface area of the unit $(N+D-1)$-sphere and $p(\bm{y})$ is the, ground truth, data distribution. However, due to rotational symmetry on the $D$-dimensional cylinder $\sum_{i=1}^D z_i^2 = r^2, \forall r > 0$, a dimensionality reduction is possible \cite{xu2023}. In fact, it suffices to track the norm of the augmented variables, $r = r(\bm{\tilde{x}}):=||\bm{z}||_2.$ For notational brevity, we redefine $\tilde{\bm{y}}:=(\bm{y},0) \in \mathbb{R}^{N+1}$ and $\tilde{\bm{x}}:=(\bm{x},r) \in \mathbb{R}^{N+1}.$ The ODE of interest is now 
\begin{equation}
    d\bm{x} = \bm{E} (\bm{\tilde{x}})_{\bm{x}} \cdot E(\bm{\tilde{x}})_r^{-1} \label{pfgmpp_ode} \, dr, 
\end{equation}
where $\bm{E}(\bm{\tilde{x}})_{\bm{x}}$, and $E(\bm{\tilde{x}})_r$, a scalar, denote the $\bm{x}$ and $r$ components of $\bm{E} (\bm{\tilde{x}}),$ respectively. Eq. \ref{pfgmpp_ode} defines a surjection between the data on the $r=0$ hyperplane and an easy-to-sample prior noise distribution on $r=r_{\max}$ hyper-cylinder \cite{xu2023}. As for diffusion models, PFGM++ employs a perturbation based objective. Let $p(r)$ the training distribution over $r$ and $p_r(\bm{x} | \bm{y})$ the perturbation kernel then the objective of interest is 
\begin{equation}
    \mathbb{E}_{r\sim p(r)} \mathbb{E}_{\bm{y}\sim p(\bm{y})} \mathbb{E}_{\bm{x} \sim p_r(\bm{x}|\bm{y})} \left\Vert f_\theta (\bm{\tilde{x}})-\frac{\bm{x}-\bm{y}}{r/\sqrt{D}} \right\Vert_2^2. 
    \label{pfgmpp_obj_final}
\end{equation}
Now, if $p_r(\bm{x}|\bm{y}) \propto 1/(|| \bm{x}-\bm{y}||_2^2+r^2)^{\frac{N+D}{2}}$ then it is possible to show that the minimizer of Eq. \eqref{pfgmpp_obj_final} is $f^*_\theta(\bm{\tilde{x}}) = \sqrt{D} \bm{E} (\bm{\tilde{x}})_{\bm{x}} \cdot E(\bm{\tilde{x}})_r^{-1}.$ As for diffusion models, a sample can be generated by solving $d\bm{x}/dr = \bm{E}(\bm{\tilde{x}})_{\bm{x}}/E(\bm{\tilde{x}})_r = f^*_\theta(\bm{\tilde{x}})/\sqrt{D}$, starting from an initial sample from the prior noise distribution, $p_{r_{\max}},$ using an iterative ODE solver. 

\subsection{Problem formulation}
We will treat the problem of obtaining a high quality reconstruction $\bm{\hat{y}} \in \mathbb{R}^N$  of $\bm{y} \in \mathbb{R}^N$ based on noisy observations $\bm{c} = \mathcal{F}(\bm{y}) \in \mathbb{R}^N$ as a statistical inverse problem, where $\mathcal{F}: \mathbb{R}^N \rightarrow \mathbb{R}^N$ is a ``catch-all'' noise degradation operator, including factors such as quantum noise \cite{chen2017}, and $N:= n\times n.$ Thus, we will assume that the data follows a prior distribution $\bm{y}\sim p(\bm{y})$ and our objective is to enable sampling from the posterior $p(\bm{y}|\bm{c})$. This approach to inverse problem solving is referred to as posterior sampling. In this paper, $\bm{y}$ will be treated as ``ground truth'' despite the fact that it may contain noise and artifacts.

\subsection{Image denoising via posterior sampling}
Our proposed method has two main components: 1) a learned PFGM++ trained in an unsupervised manner for the task of unconditional image generation. 2) a sampling scheme which regularizes the generative process, enforcing consistency with the input, condition, image. Combining the information from the prior distribution, $p(\bm{y})$, embedded in the learned empirical electric field with a modified sampling scheme enforcing data consistency allows us to sample from the desired posterior $p(\bm{y}|\bm{c}),$ providing a solution to our inverse problem. This strategy has been successfully applied in the case of diffusion models for a range of inverse problem \cite{song2022b,chung2022,chung2023,liu2023}. The key idea in this paper, is to extend this approach to the case of PFGM++. Directly extending to PFGM++ should be feasible due to the intimate connection with diffusion models. We leave theoretical analysis to future work, and instead provide empirical support that we indeed get a sample from the desired posterior. In particular, we will show that $\bm{\hat{y}} \approx \bm{y}$. As in PFGM++ \cite{xu2023}, we reuse the proposed sampling algorithm from \cite{karras2022} with an updated prior noise distribution. This is feasible due the hyperparameter translation formula, $r=\sigma \sqrt{D}$, \, $\bm{\tilde{x}}:=(\bm{x},r)$, the fact that $\sigma(t)=t$ in \cite{karras2022}, and by a change of variable $d\bm{x} = f^*_\theta(\bm{\tilde{x}})/\sqrt{D} \, dr = f^*_\theta(\bm{\tilde{x}}) \, dt$ since $dr=d\sigma \sqrt{D}=dt \sqrt{D}.$ The proposed sampling algorithm is detailed in Algorithm \ref{our_algo} with updates to sampling vs \cite{xu2023} highlighted in blue. The first update is hijacking: instead of starting from an initial sample for the prior noise distribution we hijack the reverse process at some $i=\tau \in \mathbb{Z}_+, \tau < T$ and simply inject the condition image $\bm{x}_\tau =\bm{c}.$ For the case of diffusion models, it is common to forward diffuse the condition image prior to injecting into the reverse process as in e.g., \cite{chung2022}. This will add additional stochasticity into the sampling and we did not see improvements in overall performance when taking this approach. Hence, we opted for injecting the condition image directly. In addition to hijacking, we also have a data consistency, or regularization, step. This would need to be updated for different inverse problems. For the case of image denoising, simply using the identity map will suffice. Thus, we simply mix $\bm{x}_{i+1}$ with $\bm{x}_{\tau}=\bm{c}$, where $\bm{c}$ is the condition image, with weight $w \in [0,1].$ We did consider using a low-pass filtered version of $\bm{c}$ as is done for diffusion models in e.g., \cite{chung2023}. However, this did not lead to major improvement gains and we therefore opted for this simpler formulation. We note that the proposed sampling algorithm is identical to the one used in \cite{hein2023} except for the fact that the network has now been trained using unsupervised learning and thus does no longer take the condition image $\bm{c}$ as additional input. 

\begin{algorithm}
\DontPrintSemicolon
\SetAlgoNoLine
\textcolor{blue}{Get initial data $\bm{x}_\tau=\bm{c}$}\;
\For{$i= \textcolor{blue}{\tau},...,T-1$}{
    $\bm{d}_i = (\bm{x}_i-D_\theta(\bm{x}_i,t_i))/t_i$ \; 
    $\bm{x}_{i+1}=\bm{x}_i+(t_{i+1}-t_i)\bm{d}_i$ \; 
    \If{$t_{i+1}>0$}{
        $\bm{d}_i' = (\bm{x}_{i+1}-D_\theta (\bm{x}_{i+1},t_{i+1}))/t_{i+1}$ \;
        $\bm{x}_{i+1}=\bm{x}_i+(t_{i+1}-t_i)(\frac{1}{2}\bm{d}_i+\frac{1}{2}\bm{d}_i')$ \;
    }
    $\textcolor{blue}{\bm{x}_{i+1} = w\bm{x}_{i+1}+(1-w)\bm{x}_{\tau}}$ \;
}
\Return $\bm{x}_T$
\caption{Proposed PPFM sampling. Adapted from PFGM++\cite{xu2023} with adjustments highlighted in blue.}
\label{our_algo}
\end{algorithm}

\subsection{Experiments}
\subsubsection{Datasets}
\paragraph{Mayo low-dose CT data}
For training and validation we use the publicly available dataset from the Mayo Clinic used in the American Association of Physicists in Medicine (AAPM) low-dose CT grand challenge \cite{aapm2017}. This dataset contains images from 10 patients reconstructed with medium (D30) and sharp (D45) kernels. The images are also available at 1~mm and 3~mm slice thicknesses. All images have a matrix size of $512\times512$. In this paper, we use the 1~mm slice thickness and B30 reconstruction kernel. We use the first 8 patients for training, yielding a total of 4800 slices, and the last two for validation, yielding a total of 1136 slices. We stress that, although paired normal-dose CT (NDCT) and low-dose CT (LDCT) images are available, only the NDCT images are used for training the proposed method. 

\paragraph{Photon-counting CT data}
As test data we use images of two patients from a clinical study\footnote{Swedish Ethics Review Agency 2020-04638 and 2021-01092 and prospectively consented IRB review UW-IRB: 2022-1043.} of a prototype silicon-detector-based photon-counting system developed by GE HealthCare \cite{almqvist2023}. The patients were scanned at the Karolinska Insitutet, Stockholm, Sweden (Case 1, effective diameter 28 cm, $\mathrm{CDTI_{vol}}=10.12 \; \mathrm{mGy}$) and at the University of Wisconsin–Madison, Madison, WI (Case 2, effective diameter 36~cm, $\mathrm{CDTI_{vol}}=27.64~ \mathrm{mGy}$) using scan parameters listed in table \ref{scan_table}. 70~keV virtual monoenergetic images were reconstructed with filtered backprojection on a $512 \times 512$ pixel grid with 0.42~mm slice thickness.

\begin{table*}
    \centering
    \begin{tabular}{cccc} \toprule
	   &	PCCT (Case 1) & PCCT (Case 2) \\ \midrule 
      Tube current  & 255~mA & 290~mA\\ 
      Helical pitch & 0.990:1 & 0.510:1\\ 
      Rotation time & 0.6~s & 0.7~s &\\
      kVp & 120 & 120 \\
 \bottomrule
    \end{tabular}
    \caption{Key parameters used for scanning patients on prototype photon-counting CT systems.}
    \label{scan_table} 
\end{table*}

\subsubsection{Implementation details}
Each network, $D \in \{64,128,2048\}$ and $D\rightarrow \infty$, is trained with batch size 32 using Adam \cite{kingma2017} with learning rate $2\times10^{-4}$ for 10$^5$ iterations. We use the DDPM++ architecture with channel multiplier 128, channels per resolution [1, 1, 2, 2, 2, 2, 2], and self-attention layers at resolutions 16, 8, and 4. We employ the preconditioning, exponential moving average (EMA) schedule, and non-leaky augmentation suggested in \cite{karras2022} with an augmentation probability of $15\%$. In addition, to further prevent overfitting, we set dropout probability to $10\%$. To enable efficient training, the network is trained on randomly extracted $256 \times 256$ patches. In other words, the unconditional generator is not trained to generate full resolution CT images but rather patches extracted from CT images. In addition to facilitating efficient training, this serves to further augment our dataset and therefore help prevent overfitting. In order to reduce graphics memory requirements even further, we train using mixed precision. We note that the same configuration but with $512\times512$ images would exceed the memory available on a NVIDIA A6000 48~GB GPU. To achieve NFE=1, we fix $\tau:=T-1.$ We set $T$ and $w$, crucial hyperparameters in Algorithm \ref{our_algo}, by grid search over $T\in \{4,8,16,32,64\}$ and $w\in \{0.5,0.6,0.7,0.8,0.9,1.0\}$ using Learned Perceptual Image Patch Similarity (LPIPS) \cite{zhang2018} on the validation set as selection criteria. Lowest (best) LPIPS was achieved for $T=8$ and $w=0.5$. Since $w=0.5$ is a corner case, we also tried $w=0.4$ but this did not improve performance. We refer to the proposed method as ``single-step'' despite including a second step, regularization, since this the time required from this operation is negligible and NFE=1. 

\subsubsection{Comparison to other methods}
We compare our results with popular non-deep learning-based, supervised, and unsupervised image denoising techniques. For non-deep learning-based image denoising, we chose a version of BM3D \cite{makinen2020} since this was shown to be the best performing method in the non-deep learning category for low-dose CT image denoising in \cite{chen2017}. We use bm3d.py\footnote{\url{https://pypi.org/project/bm3d/}.} and set the parameter $\sigma_{\text{BM3D}}$ by measuring the standard deviation in a flat region-of-interest (ROI) in the low-dose CT validation data. For supervised techniques we use WGAN-VGG \cite{yang2018}, consistency models \cite{song2023} and PPFM \cite{hein2023}. The WGAN-VGG was trained on randomly extracted $64\times 64$ from the low-dose CT training data with hyperparameters as specified in \cite{yang2018}. Consistency models \cite{song2023} are the current state-of-the-art diffusion style model with NFE=1. However, consistency models were developed for the task of unconditional image generation. To the best of our knowledge, \cite{hein2023} is the first implementation of a consistency model for conditional (image-to-image) generation. It is reasonable to surmise that the ``trick'' of feeding the condition image as additional input to the network to directly learn a trajectory to the posterior distribution of interest, a technique successfully used for diffusion models \cite{batzolis2021,saharia2022b,saharia2023} and PFGM++ \cite{hein2023}, will also work for consistency distillation. This hypothesis was supported empirically in \cite{hein2023}. Minimal adjustments were made to the official implementation\footnote{\url{https://github.com/openai/consistency_models}.} to feed the condition images $\bm{c}$ as additional input. The network is trained on randomly extracted $256\times 256$ patches from the low-dose CT data. Random rotations and mirrors are applied as data augmentation. Hyperparameters for sampling and training are set as for the LSUN $256\times 256$ experiments\footnote{As specified in \url{https://github.com/openai/consistency_models/blob/main/scripts/launch.sh}.}, except for batch size with had to be reduced to 4 to fit in the memory of a single NVIDIA A6000 48~GB GPU. An EDM is first trained for $3\times10^5$ iterations and then distilled into a consistency model for $6\times10^5$ iterations. We will refer to this model as CD (consistency distillation). Training and sampling for PPFM are as specified in \cite{hein2023}. We use the $D=64$ case since this was the best performing. The results for BM3D, WGAN-VGG, CD, and PPFM are derived directly from reference \cite{hein2023}. Finally, as an example of an unsupervised method we use Noise2Void \cite{krull2019}. We employ the same hyperparameters as for the BSD68 dataset in the original paper using code from the official repository\footnote{\url{https://github.com/juglab/n2v}.}.

\subsection{Evaluation methods}
For quantitative image quality assessment we employ structural similarity index (SSIM \cite{wang2004}), peak signal-to-noise ratio (PSNR), and LPIPS \cite{zhang2018}. SSIM and PSNR are very well established metrics in the medical imaging literature, but these relatively simple metrics do not necessarily correspond well to human perception \cite{zhang2018}. For instance, PSNR is inverse proportional to the $\ell_2$ Euclidean distance and this simple pixel-wise metrics correlate well with human perception. This is particularly evident for the case of over-smoothing. LPIPS, similar to perceptual loss functions, uses features extracted from the images using pretrained convolutional neural networks (CNNs).  Over an extensive series of experiments, reference \cite{zhang2018} demonstrates that LPIPS correlates better with human preferences than simple metrics such as SSIM and PSNR. We use the AlexNet \cite{krizhevsky2014} version of LPIPS from the official implementation\footnote{\url{https://github.com/richzhang/PerceptualSimilarity}.}.

\section{Results}\label{sec3}
\begin{figure}
    \centering
    \includegraphics[width=\columnwidth]{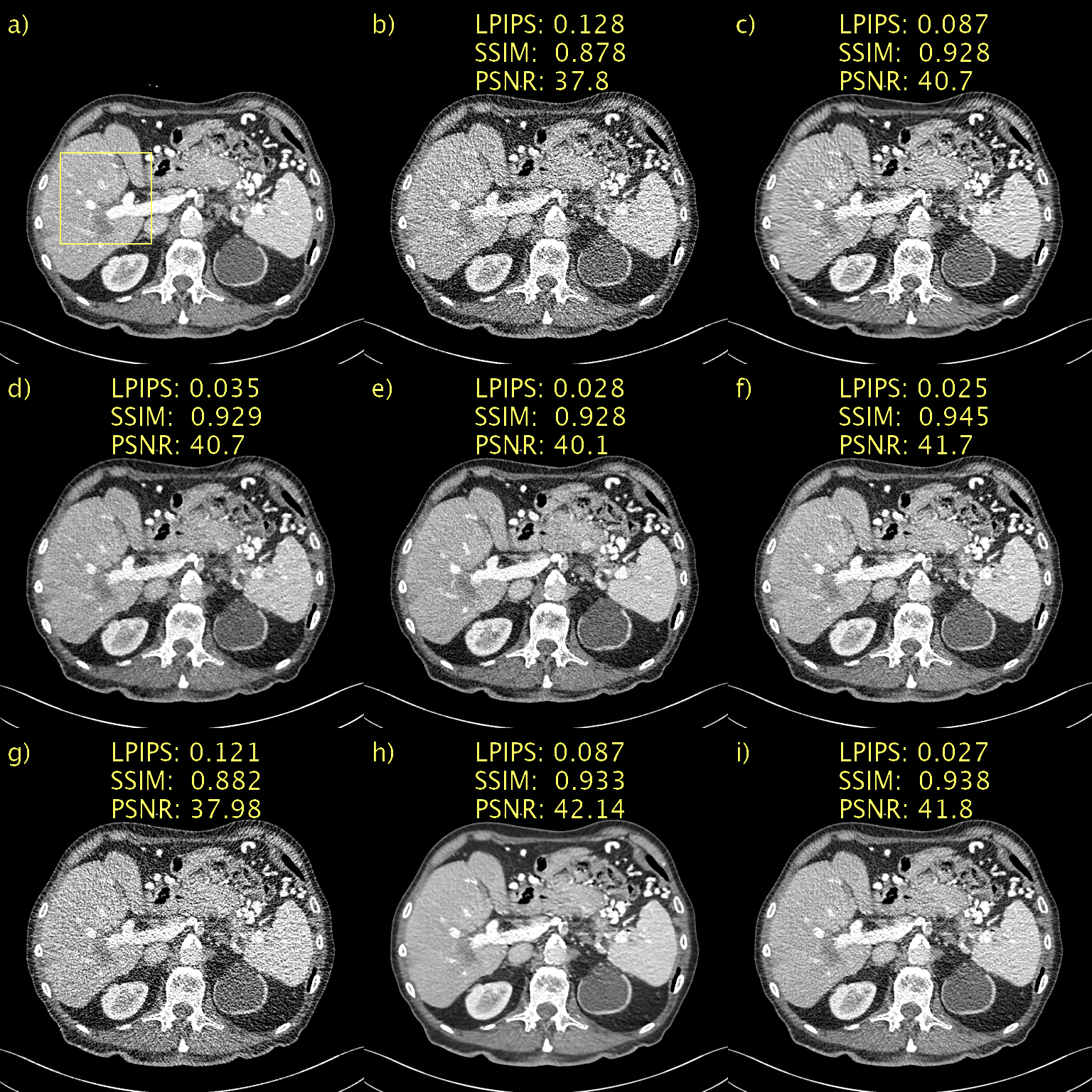}
    \caption{Results on the Mayo low-dose CT dataset. Abdomen image with a metastasis in the liver. a) NDCT, b) LDCT, c) BM3D, d) WGAN-VGG, e) CD, f) PPFM, g) Noise2Void h) $D\rightarrow \infty$ i) $D=128$. Yellow box indicating ROI shown in Fig. \ref{1049_ex}. 1~mm-slices. Window setting [-160,240]~HU.}
    \label{1049}
\end{figure}

\begin{figure}
    \centering
    \includegraphics[width=\columnwidth]{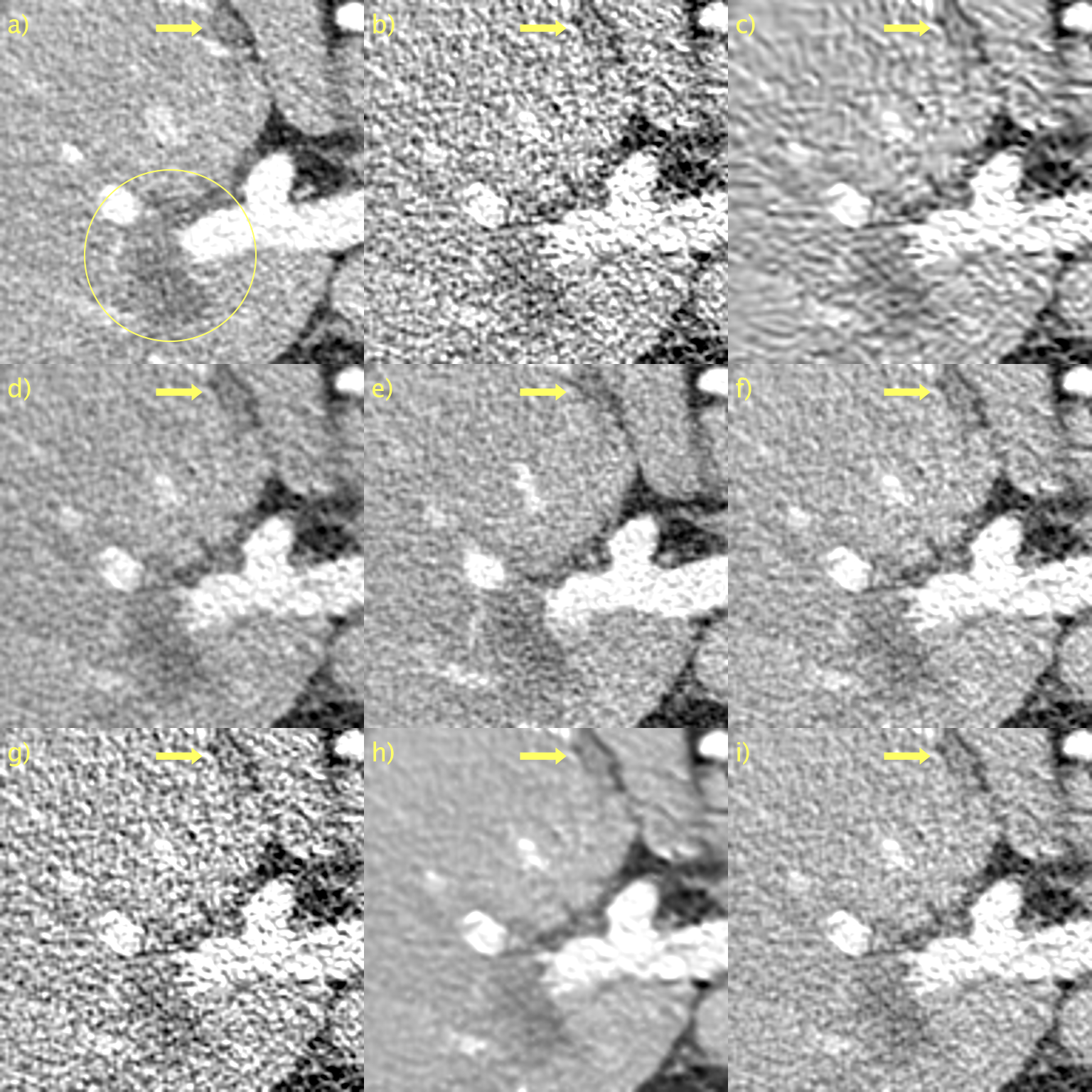}
    \caption{ROI in Fig. \ref{1049} magnified to emphasize details. a) NDCT, b) LDCT, c) BM3D, d) WGAN-VGG, e) CD, f) PPFM, g) Noise2Void h) $D\rightarrow \infty$ i) $D=128$. Yellow circle added to emphasize lesion. 1~mm-slices. Window setting [-160,240]~HU.}
    \label{1049_ex}
\end{figure}

Quantitative results, the mean and standard deviation of LPIPS \cite{zhang2018}, SSIM \cite{wang2004}, and PSNR for the low-dose CT validation set are available in Table \ref{table_quant}. The overall top performer is PPFM. However, as mentioned above, SSIM and PSNR do not necessarily correspond closely to human perception. As expected, our proposed method is bounded in performance by PPFM since it is unsupervised. Moreover, we can see that the proposed method, for $D=128$ and $D=64$, is actually better in terms of LPIPS than WGAN-VGG, a supervised method. Compared to CD, the state-of-the-art diffusion-style model with NFE=1, the performance of the proposed method is slightly worse. However, importantly, CD was trained in a supervised manner whereas proposed is unsupervised. Keeping this in mind, the proposed method with $D=128$ actually performs very competitively. Comparing the two unsupervised methods, Noise2Void and the proposed, we can see that the latter performs very favorably. Noise2Void yields only a very marginal improvement over the LDCT images. 
\begin{table}
    \centering
    \begin{tabular}{cccc} \toprule
	& LPIPS ($\downarrow$)& SSIM ($\uparrow$)&	PSNR ($\uparrow$)\\ \midrule 
        LDCT   & $0.075 \pm 0.02$ & $0.94 \pm 0.02 $ & $41.5\pm 1.6$ \\
        BM3D\cite{makinen2020} & $0.050 \pm 0.01$ & $\bm{0.97} \pm 0.01 $ & $45.0\pm 1.6$\\  
        WGAN-VGG\cite{yang2018} & $0.019 \pm 0.01$ & $0.96 \pm 0.01 $ & $43.2\pm 0.9$\\
        CD\cite{song2023} & $0.013 \pm 0.00$ & $0.96 \pm 0.01 $ & $43.1\pm 1.0$\\
        PPFM\cite{hein2023} & $\bm{0.010} \pm 0.00$ & $\bm{0.97} \pm 0.01 $ & 
        $\bm{45.4} \pm 1.4$\\ 
        Noise2Void\cite{krull2019} & $0.069 \pm 0.02$ & $0.94 \pm 0.02 $ & $41.7\pm 1.6$ \\   
        \midrule
        Proposed & & & \\
        \midrule 
        $D\rightarrow \infty$ & $0.059 \pm 0.02$ & $0.96 \pm 0.01 $ & $44.8\pm 0.9$\\
        $D=2048$ & $0.058 \pm 0.02$ & $0.96 \pm 0.01 $ & $44.9\pm 0.9$\\ 
        $D=128$ & $0.014 \pm 0.00$ & $\bm{0.97} \pm 0.01 $ & $45.3\pm 1.4$\\
        $D=64$ & $0.015 \pm 0.00$ & $\bm{0.97} \pm 0.01 $ & $\bm{45.4}\pm 1.4$\\ 
    \bottomrule \\
    \end{tabular}
    \caption{Mean and standard deviation of LPIPS, SSIM, and PSNR in the low-dose CT validation set. WGAN-VGG, CD, and PPFM are supervised methods. Noise2Void and proposed are unsupervised. BM3D is non-deep learning. $\downarrow$ means lower is better. $\uparrow$ means higher is better. Best results in bold.}
    \label{table_quant} 
\end{table}
Qualitative results, alongside LPIPS, SSIM, and PSNR, for a representative slice from the Mayo low-dose CT validation data appear in Fig. \ref{1049}. For the sake of brevity, we show results for the proposed method only for $D\rightarrow \infty$ and, best performing, $D=128.$ The former is included as an interesting case since it corresponds to a diffusion model, instead of PFGM++. This patient has a metastasis in the liver and we have included a 
magnification of this region of interest (ROI) in Fig. \ref{1049_ex}. a) and b) show the NDCT and LDCT images, respectively for reference. BM3D, shown in c), performs well in terms of noise suppression whilst preserving salient details. Nevertheless, this comes at a cost of artifacts that make the image appear a somewhat smudgy. WGAN-VGG, CD, and PPFM, all supervised methods, shown in d), e) and f), all suppress noise effectively and keep key details intact. We have overlaid a yellow arrow to point out a detail that appears in CD, shown e), but in none of the other images, including NDCT and LDCT. It appears that CD has added a feature into the image that appears realistic, but which we know is not genuine, given that we have the LDCT and NDCT images as reference. Such inaccurate removal or addition of details is loosely referred to as hallucination \cite{koetzier2023}. Noise2Void, shown in g), seems to in essence reproduce the LDCT image. However, quantitatively there is a marginal improvement. However, as shown in i), the proposed method (with $D=128$) effectively suppresses the noise whilst keeping salient features intact. Qualitatively, it is difficult to discriminate between the proposed method with $D=128,$ which is an unsupervised method, and PPFM, shown in f), a supervised method. Quantitatively for this particular slice, we can see that PPFM performs slightly better. Comparing $D\rightarrow \infty$, shown in h), with $D=128$ in i) demonstrates the performance gains afforded by the PFGM++ framework. In particular, the proposed method with $D\rightarrow \infty$ appears over-smoothed and somewhat blurred. 

\begin{figure}
    \centering
    \includegraphics[width=\columnwidth]{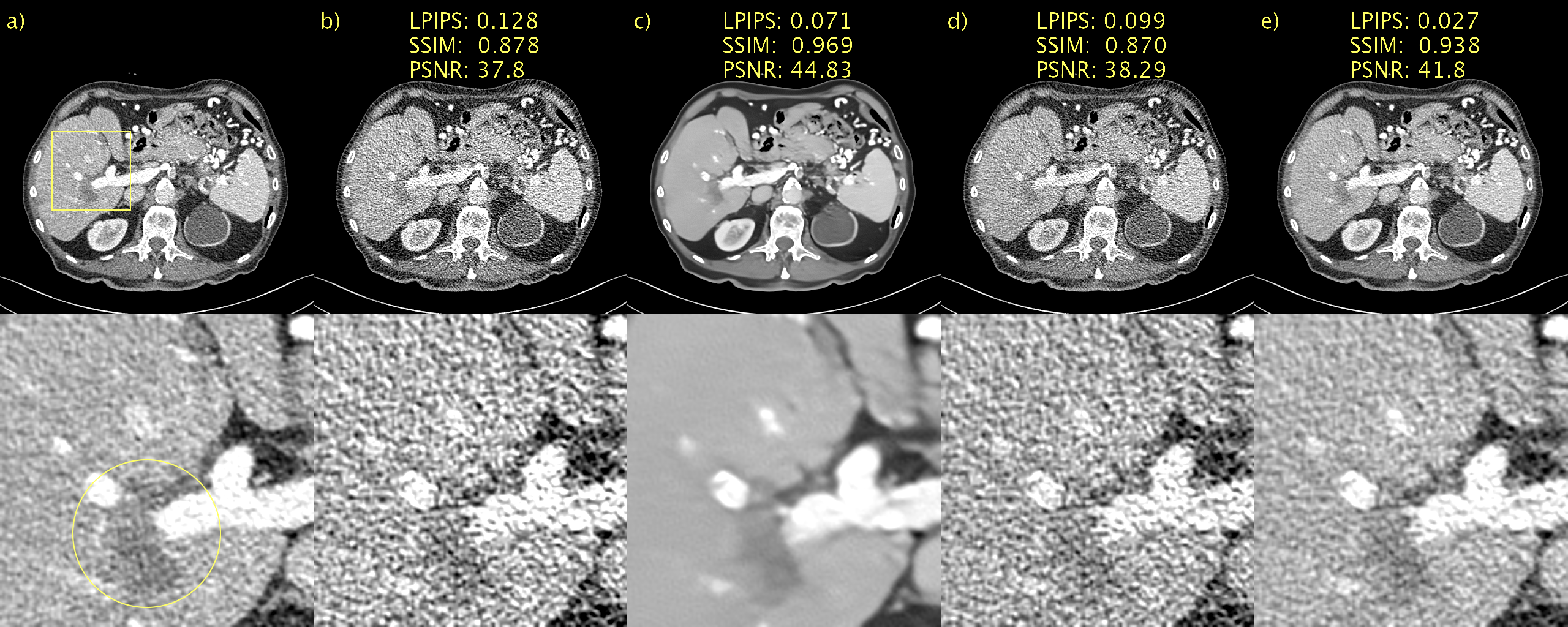}
    \caption{Ablation study of regularizer used in our proposed sampler. a) NDCT, b) LDCT, c) only hijacking, d) only regularization, e) hijacking and regularization. Yellow circle added to emphasize lesion. 1~mm-slices. Window setting [-160,240]~HU.}
    \label{1049_ablation}
\end{figure}

Results from an ablation study of the proposed sampler are available in Fig. \ref{1049_ablation}. a) and b) show the NDCT and LDCT images for easy reference. In c) we hijack but omit regularization ($\tau=T-1,w=1$). The result is an image that has been very aggressively denoised. This is a direct consequence of the small $T$ and corresponding large step-size. This represents a further clean demonstration of how SSIM and PSNR fail to adequately penalize blurring as c) appears blurry to a human observed yet performs very well according to SSIM and PSNR. LPIPS, on the hand, penalizes this quite heavily. In d), we regularize but employ no hijacking ($\tau=0,w=0.5$). In this case we start with an initial sample from the prior noise distribution. The amount of regularization in this setting seems excessive and we recover an image where $\bm{\hat{y}}\approx \bm{c}.$ Finally, in e) we employ hijacking and regularization ($\tau=T-1,w=0.5$), resulting in a very pleasing image where $\bm{\hat{y}}\approx \bm{y}.$ Consequently, there is a significant reduction (improvement) in LPIPS. 

\begin{figure}
    \centering
    \includegraphics[width=\columnwidth]{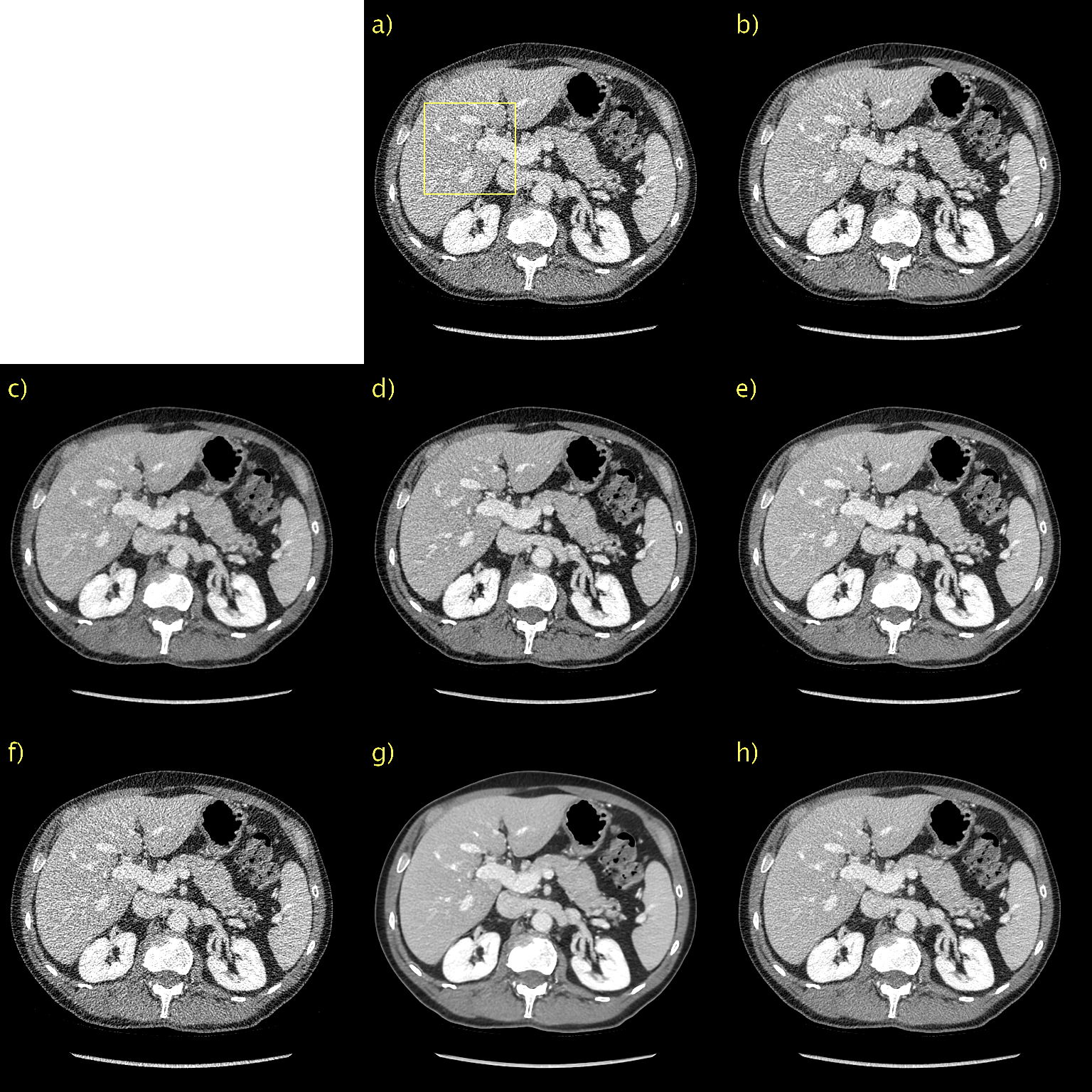}
    \caption{Results for first PCCT test case. a) Unprocessed, b) BM3D, c) WGAN-VGG, d) CD, e) PPFM, f) Noise2Void g) $D\rightarrow \infty$ h) $D=128$. No ground truth available. Yellow box indicating ROI shown in Fig. \ref{233_ex}. 0.42~mm-slices. Window setting [-160,240]~HU.}
    \label{233}
\end{figure}

\begin{figure}
    \centering
    \includegraphics[width=\columnwidth]{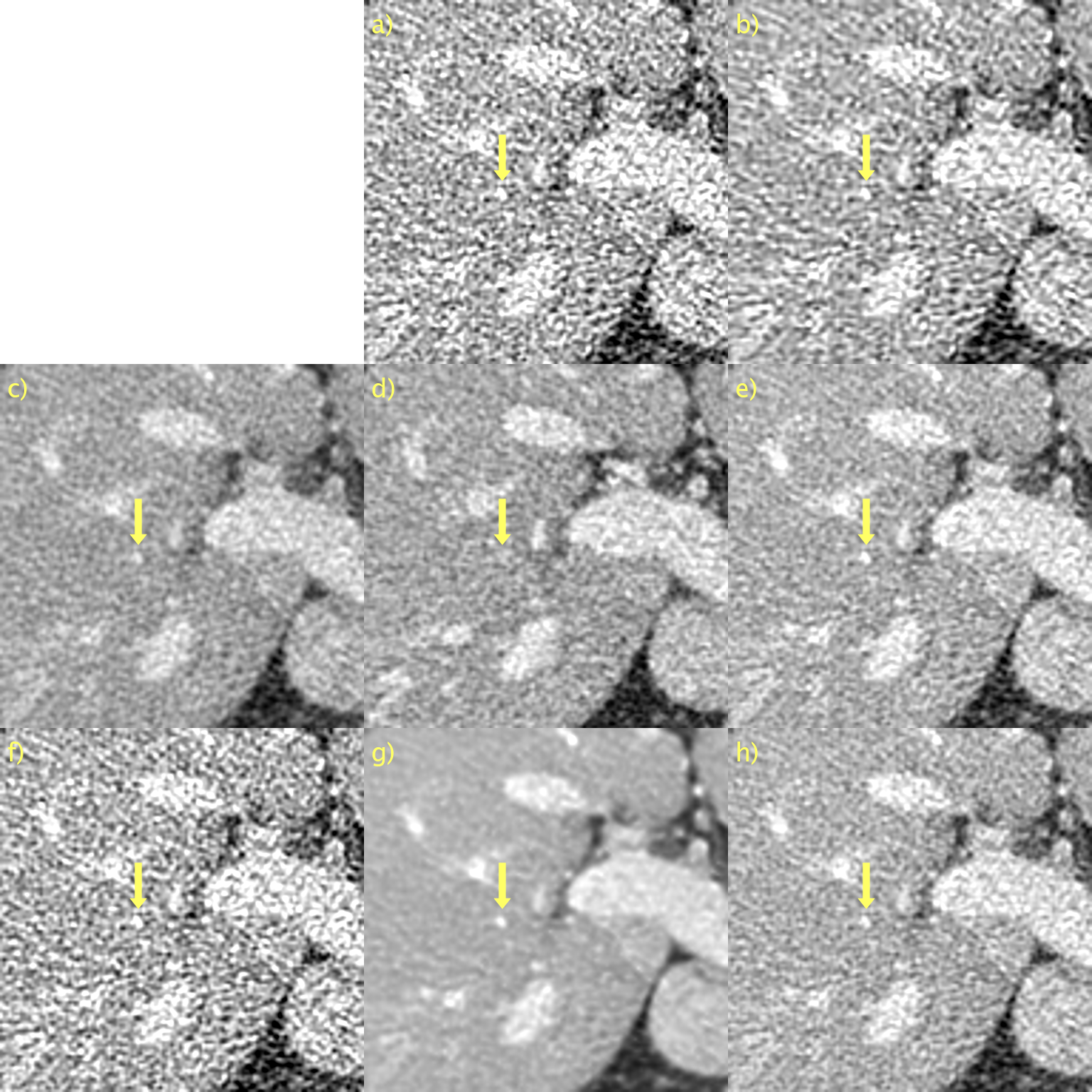}
    \caption{ROI in Fig. \ref{233} magnified to emphasize details. a) Unprocessed, b) BM3D, c) WGAN-VGG, d) CD, e) PPFM, f) Noise2Void g) $D\rightarrow \infty$ h) $D=128$. No ground truth available. Yellow arrow placed to emphasize detail. 0.42~mm-slices. Window setting [-160,240]~HU.}
    \label{233_ex}
\end{figure}

Qualitative results for a representative slice in the first PCCT test case appear in Fig. \ref{233} with a magnification of the indicated ROI in Fig. \ref{233_ex}. Due to the lack of ``ground truth'' images, we resort to qualitative evaluation. It is also difficult to discriminate signal from noise when it comes to fairly small and low-contrast details. We simply define good performance as accurately reproducing the unprocessed image, shown in a), but with a lower noise level. BM3D, in b), seems to generalize poorly from the low-dose CT data. This is likely due to difference in noise characteristics; one would need to re-estimate $\sigma_{\text{BM3D}}.$ All other methods seem to generalize well in the sense that there are no major changes in performance. There is a significant performance gain for proposed method with $D=128$ compared to $D\rightarrow \infty.$ This is most visible in the magnified ROI in Fig. \ref{317_ex} as $D\rightarrow \infty$, shown in g), is significantly more blurry than $D=128,$ shown in h). We have overlaid a yellow arrow, indicating a detail of interest. Since we do not have a ground truth, and we are considering a single slice, we cannot definitively state that this is not just a noise spike. However, since it is clearly visible in the unprocessed image, we want it to be visible in the processed images as well. As we can see, this is indeed the case with the notable exception of CD, shown in d). Moreover, the contrast of this detail appears to vary, and seems to be much more well defined for the proposed with $D=128$ than for WGAN-VGG, shown in c). Hence, it appears than the proposed method, despite being unsupervised, is able to perform very competitively even compared to supervised methods such as WGAN-VGG and CD. 

\begin{figure}
    \centering
    \includegraphics[width=\columnwidth]{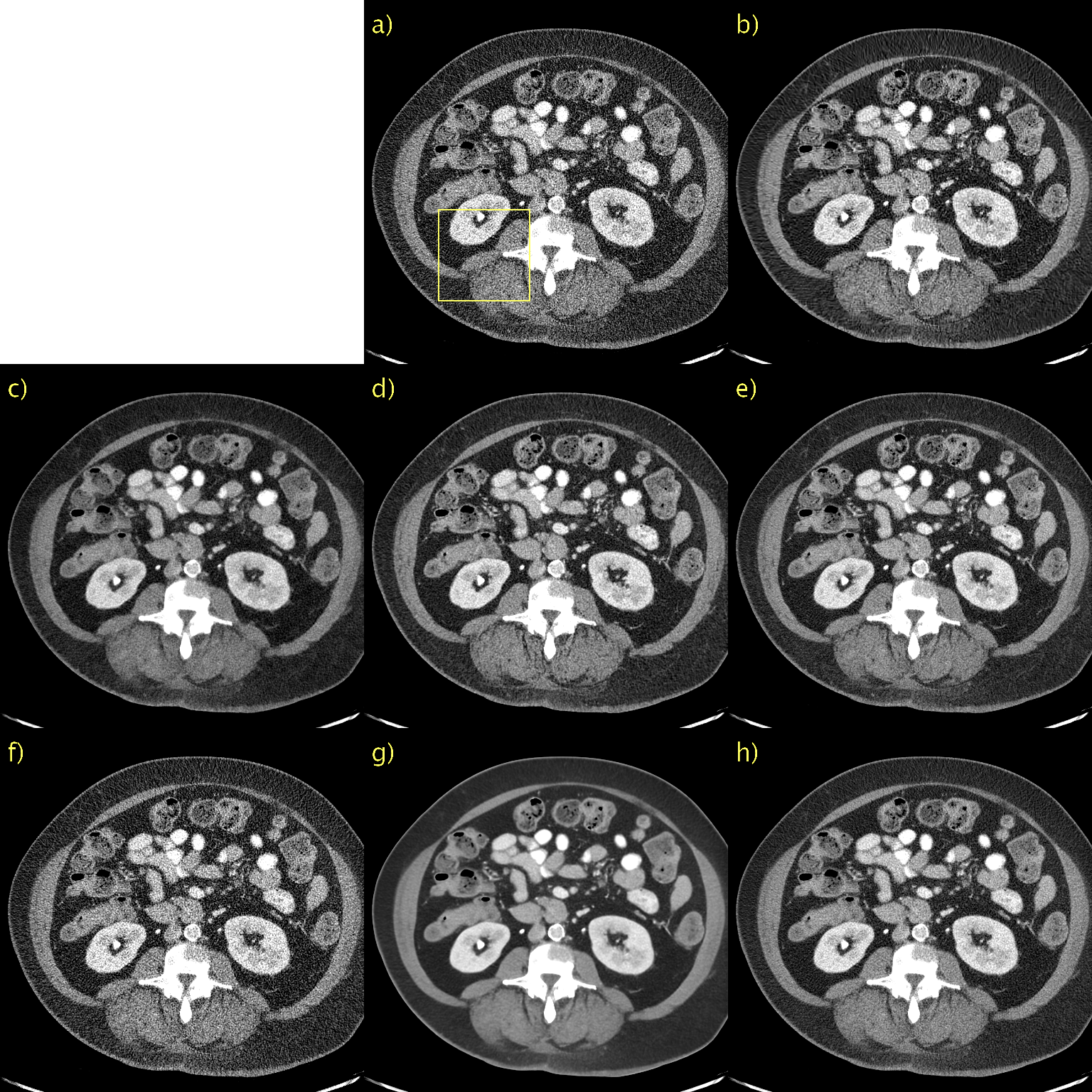}
    \caption{Results for second PCCT test case. a) Unprocessed, b) BM3D, c) WGAN-VGG, d) CD, e) PPFM, f) Noise2Void g) $D\rightarrow \infty$ h) $D=128$. No ground truth available. Yellow box indicating ROI shown in Fig. \ref{317_ex}. 0.42~mm-slices. Window setting [-160,240]~HU.}
    \label{317}
\end{figure}

\begin{figure}
    \centering
    \includegraphics[width=\columnwidth]{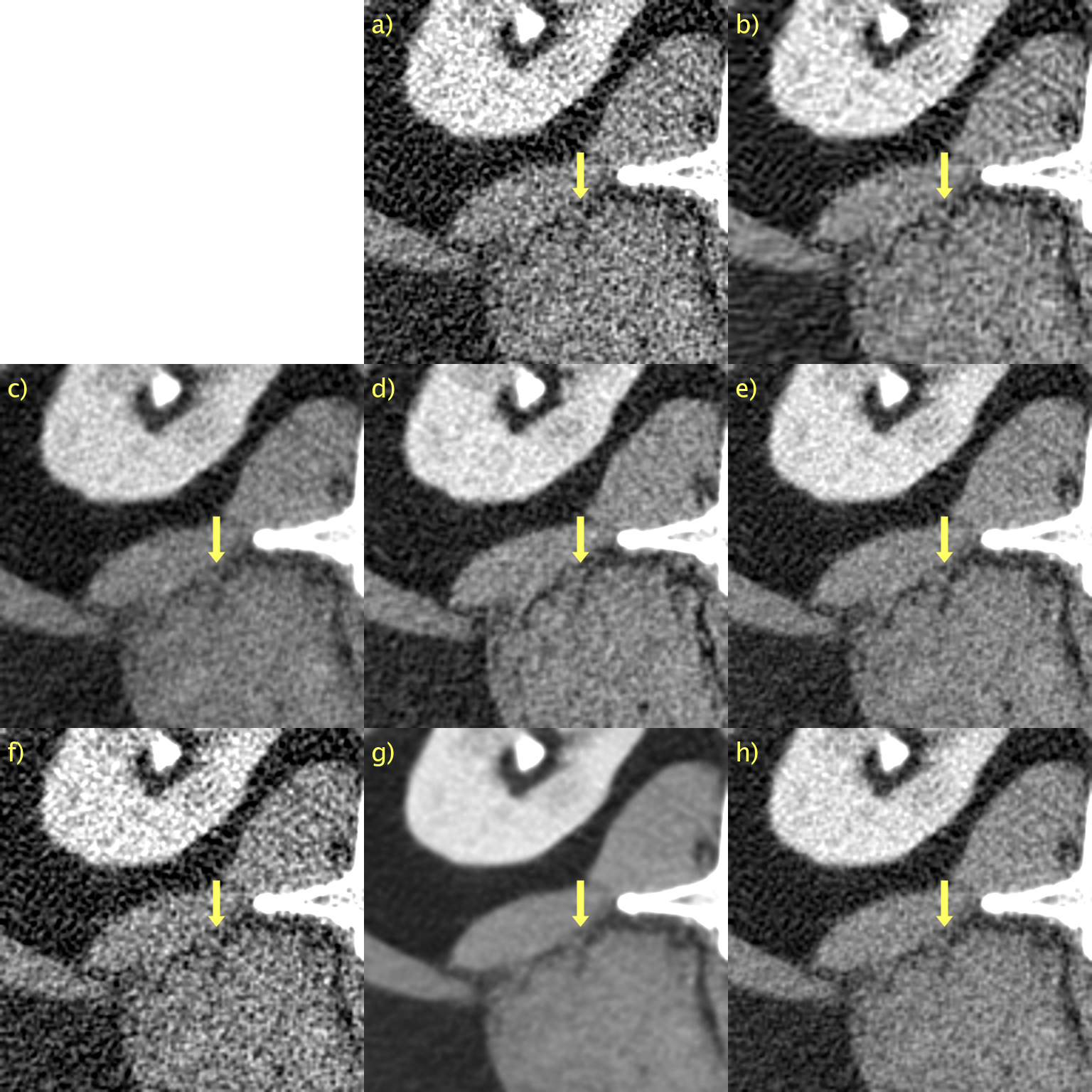}
    \caption{ROI in Fig. \ref{317} magnified to emphasize details. a) Unprocessed, b) BM3D, c) WGAN-VGG, d) CD, e) PPFM, f) Noise2Void g) $D\rightarrow \infty$ h) $D=128$. No ground truth available. Yellow arrow placed to emphasize detail. 0.42~mm-slices. Window setting [-160,240]~HU.}
    \label{317_ex}
\end{figure}

Results for a representative slice in the second PCCT test are shown in Fig. \ref{317} and Fig. \ref{317_ex}. Again, no ``ground truth'' is available. BM3D, in b), seems to perform better than for the first PCCT test case. The denoising performance is now more aligned with what was observed for the low-dose CT validation data. The difference in performance is likely due to differences in noise characteristics and lack of generalization. Lack of generalization does not seem to be an issue for any other method as performance is very consistent across the validation and test data. We have again placed a yellow arrow to indicate details of interest, in this case fat in the back muscle. Contrast is difficult to assess qualitatively when comparing proposed with $D\rightarrow \infty$, shown in g), and $D=128$, shown in h), due to the large difference in noise level. $D\rightarrow \infty$ is definitely over-smoothed and thus more blurry; however, the contrast of this particular detail seem to be fairly well preserved. The proposed method with $D=128$ again performs very competitively compared to WGAN-VGG, shown in c), as can be seen when considering the contrast of fat and muscle.

\section{Discussion}\label{sec5}
Since we are interested in image denoising we were able to choose the simplest possible data consistency (or regularization) step, the identity map. One interesting direction of future research is to see how well the proposed method generalizes to other inverse problems. We surmise that it is sufficient to update only the regularization step, as is the case for similar techniques based on diffusion models \cite{song2022b,chung2023}. One interesting application is to combine image denoising and super-resolution. However, it is possible that one also needs to update the hijacking method as it may be the case that injecting the condition image directly into the sampling process will fail to generalize beyond the problem of image denoising. Even within the two tasks of denoising low-dose and photon-counting CT images, noise characteristics vary widely. In particular, CT images are routinely reconstructed using different kernels, slice thicknesses, fields-of-view and matrix sizes. These factors may result in reconstructed images with vastly different noise characteristics. As shown in \cite{huber2021}, this may adversely impact the performance of image denoising techniques. Hence, it is possible that one needs to update the hyperparameters in the sampling algorithm, including the consistency step, to attain good performance over a wide range of settings. 

Another interesting avenue of future research is to extend the proposed method to 3D denoising. Given the structure of CT data, 2D denoising is discarding an abundance of rich information  by not considering adjacent slices. It is possible that using this additional information can, for instance, aid in recovering more of the details observed in the NDCT image from the low-dose CT data. In particular, using information from adjacent slices may help to better differentiate noise from signal. Interestingly, extending the proposed method to 3D denoising can be effected in two main ways. First, we can simply retrain the network using 3D data. The benefit of this approach is that this will allow the network to optimize the use of information from adjacent slices. However, the downside is that this  requires retraining. Another possibility is to update only the data consistency step. One can keep the learned prior from 2D data, thus not requiring any retraining, and combine it with a regularization step that utilizes information from adjacent slices. Finally, one could combine these two aforementioned approaches. We plan to extend the proposed method to 3D denoising in future work. 

For current generation photon-counting CT scanners we are usually interested in spectral (material-resolved) CT. Hence, it of interest to extend this method to the spectral case. Similar to the 3D case, one could allow the network to accept more channels, where the channels could be material basis images or virtual monoenergetic images at two or more energy levels. Another possibility is to simply train two separate networks for two different material basis images or virtual monoenergetic images at different energy levels. 

\section{Conclusion}\label{sec6}
In practice, paired data are usually not available for CT image denoising. In this paper, we propose an unsupervised version of PPFM \cite{hein2023} and demonstrate that despite imposing a significantly laxer data requirement, there is only a small drop in overall performance. To achieve this, we combine a PFGM++ \cite{xu2023} trained in an unsupervised manner for the task of unconditional generation, with a sampling scheme which enforces consistency with the input, or condition image, to enable sampling from the desired posterior. Our proposed method includes the corresponding method based on diffusion models (EDM \cite{karras2022}) and we demonstrate that the PFGM++, with $D$ as an additional hyperparameter, yields significant performance gains. Our results indicate competitive performance compared to popular supervised, unsupervised, and non-deep learning-based image denoising techniques, including state-of-the-art diffusion-style models with NFE=1, consistency models, on clinical low-dose CT data and clinical images from a prototype PCCT system.

\bibliography{sn-bibliography}

\end{document}